\documentclass[a4paper, 11pt, oneside]{Thesis}
\graphicspath{{Figures/}}

\usepackage[square, numbers, comma, sort&compress]{natbib}

\usepackage{verbatim}
\usepackage{vector}
\hypersetup{urlcolor=blue, colorlinks=false}

\begin{document}
\frontmatter

\title      {Single-shot Quantum State Merging}
\authors    {\texorpdfstring
            {\href{Mario.Berta@physik.uni-muenchen.de}{Mario Berta}}
            {Mario Berta}
            }
\addresses  {\groupname\\\deptname\\\univname}

\date       {February 2008 (Updated in \today)}
\subject    {}
\keywords   {}

\maketitle

\fancyhead{}
\rhead{\thepage}
\lhead{}

\pagestyle{fancy}

\pagestyle{empty}

\addtotoc{Abstract}
\abstract{
\addtocontents{toc}{\vspace{1em}}

We consider an unknown quantum state shared between two parties, Alice and Bob, and ask how much quantum communication is needed to transfer the full state to Bob. This problem is known as state merging and was introduced in [Horodecki et al., Nature, 436, 673 (2005)]. It has been shown that for free classical communication the minimal number of quantum bits that need to be sent from Alice to Bob is given by the conditional von Neumann entropy. However this result only holds asymptotically (in the sense that Alice and Bob share initially many identical copies of the state) and it was unclear how much quantum communication is necessary to merge a single copy. We show that the minimal amount of quantum communication needed to achieve this single-shot state merging is given by minus the smooth conditional min-entropy of Alice conditioned on the environment. This gives an operational meaning to the smooth conditional min-entropy.

}

\clearpage

\acknowledgements{
\addtocontents{toc}{\vspace{1em}}

I would like to thank Renato Renner for his instructive and friendly supervision. I particularly appreciated that he always had time to discuss my problems. I also wish to thank all the other members of the Quantum Information Science group at the ETH Zurich. I have always enjoyed the pleasant atmosphere of this research group. I would especially like to express my thanks to Roger Colbeck and Marco Tomamichel for many stimulating and fruitful scientific and non-scientific discussions. I am also grateful to J\"urg Wullschleger, Matthias Christandl, Nilanjana Datta, Francesco Buscemi, Nicolas Dutil and Stefan Hengl for pointing out errors in previous versions of this work.

}
\clearpage

\pagestyle{fancy}

\lhead{\emph{Contents}}
\tableofcontents

\addtocontents{toc}{\vspace{2em}}

\mainmatter
\pagestyle{fancy}

\newcommand{\bra}[1]{\langle #1|}
\newcommand{\ket}[1]{|#1\rangle}
\newcommand{\braket}[2]{\langle #1|#2\rangle}
\newcommand{\id}{\mathrm{id}}

\chapter{Introduction}
\label{intro}
\lhead{Chapter 1. \emph{Introduction}}

The concept of quantum state merging was introduced by Horodecki et al. in 2005~\cite{horodecki:05} (for a more detailed discussion see~\cite{horodecki:07}). They consider a quantum information source $\rho_{AB}$ that emits a sequence of unknown quantum states $\ket{\psi^{1}}_{AB},\ket{\psi^{2}}_{AB},\ldots$ where the $A$-part of this goes to Alice and the $B$-part to Bob. Then they ask how much quantum communication is needed on average to bring the full states to Bob if one allows classical communication for free. It turns out that the minimal rate of quantum communication is given by the conditional von Neumann entropy $S(A\vert B)=S(AB)-S(B)$.

However the results of Horodecki et al.~are only asymptotic results in the sense that the conditional von Neumann entropy only quantifies how much quantum communication is needed on average. Especially they do not tell us how much quantum communication is needed to transfer one particular state.

We analyze this single-shot case and allow an error $\epsilon$ in the state transfer. Our main result is that the minimal quantum communication needed for $\epsilon$-error single-shot state merging is basically equal to minus the $\epsilon$-smooth conditional min-entropy of Alice conditioned on the environment $R$ (cp. Chapter 4 for a precise definition of $R$). Because the smooth conditional min-entropy asymptotically converges to the conditional von Neumann entropy, we can reproduce the results of Horodecki et al.

This thesis is organized as follows. We start with stating some basic facts about quantum information theory in Chapter 2. Then in Chapter 3 (smooth) min- and max-entropy are introduced and some of its properties are discussed. In Chapter 4 we give the precise definition of $\epsilon$-error single-shot state merging and proof the main result rigorously. Finally we discuss the results in Chapter 5.

\chapter{Preliminaries}
\label{pre}
\lhead{Chapter 2. \emph{Preliminaries}}

In this chapter we review some basic facts about quantum information theory to present our notation and choice of definitions. Note that we make no claim to be complete.

\section{Postulates of Quantum Mechanics}

In this thesis we assume that all Hilbert spaces are finite dimensional. Although some statements also hold for infinite dimensional Hilbert spaces, the proofs of the main results do not.

\begin{itemize}
\item A state of a quantum mechanical system with $d$ degrees of freedom can be represented by a normalized nonnegative linear operator $\rho$ on a $d$-dimensional Hilbert space $\mathcal{H}$, where normalization is with respect to the trace norm: $\left\|\rho\right\|_{1}=\textrm{tr}(\rho)=1$ (cp.~Section 2.2). In the following these operators are called density matrices and we denote the set of density matrices on $\mathcal{H}$ by $\mathcal{B}(\mathcal{H})$ . A density matrix $\rho\in\mathcal{B}(\mathcal{H})$ is called pure iff the dimension of the support of $\rho$ is equal to one, i.e. $\rho=\ket{\psi}\bra{\psi}$ for some $\ket{\psi}\in\mathcal{H}$.

\item The evolution of a closed quantum mechanical system is described by a unitary transformation $U$, i.e. $\rho'=U\rho U^{\dagger}$.

\item A quantum measurement is described by a collection $\left\{M_{x}\right\}_{x\in X}$ of measurement operators that satisfy $\sum_{x}M_{x}^{\dagger}M_{x}=\id$. The probability that an outcome $x$ occurs is $\textrm{tr}(M_{x}\rho M_{x}^{\dagger})$ and the post-measurement state is then $\rho_{x}=\frac{M_{x}\rho M_{x}^{\dagger}}{\textrm{tr}(M_{x}\rho M_{x}^{\dagger})}$. If one is ignorant of the measurement outcome, the post-measurement state is given by $\rho'=\sum_{x}M_{x}\rho M_{x}^{\dagger}$. A measurement is called projective iff the measurement operators $M_{x}$ are orthogonal projectors.

\item The Hilbert space of a composite system is the tensor product of the Hilbert spaces of the individual systems.
\end{itemize}

The evolution of quantum states can equivalently be described with quantum operations. A quantum operation is a completely positive and trace preserving (CPTP) map from the set of density matrices on a input Hilbert space $\mathcal{H}$ to the set of density matrices on a output Hilbert space $\mathcal{H}'$. It can be shown that every CPTP map can be written in the form
\begin{equation}
\Lambda(\rho)=\sum_{k}E_{k}\rho E_{k}^{\dagger}\ ,
\end{equation}
where the $E_{k}$ are linear operators from $\mathcal{H}$ to $\mathcal{H}'$ that satisfy the completeness relation $\sum_{k}E_{k}^{\dagger}E_{k}=\id$. It also holds the converse, that every map of this form is a CPTP map. For proofs, see~\cite{nielsen:00} pages 367-370.

\section{Distance Measures}

How close are two states $\rho,\sigma\in\mathcal{B}(\mathcal{H})$? Motivated by this question we introduce two distance measures in this section. We start with giving two norms on the vector space of linear operators on a Hilbert space.

\begin{definition}
Let $\rho$ be a linear operator on a Hilbert space $\mathcal{H}$. The trace norm of $\rho$ is defined by $\left\|\rho\right\|_{1}=\textrm{tr}(\sqrt{\rho^{\dagger}\rho})$ and the Hilbert-Schmidt norm is defined by $\left\|\rho\right\|_{2}=\sqrt{\textrm{tr}(\rho^{\dagger}\rho)}$.
\end{definition}

The metric induced by the trace norm is called trace distance and is a measure of closeness for quantum states. It turns out that applying a quantum operation can never increase the trace distance.

\begin{lemma}
Let $\rho$, $\sigma\in\mathcal{B}(\mathcal{H})$ and let $\Lambda$ be a CPTP map. Then
\begin{equation}
\left\|\rho-\sigma\right\|_{1}\geq\left\|\Lambda(\rho)-\Lambda(\sigma)\right\|_{1}\ .
\end{equation}
In addition, if $\Lambda$ is an isometry then the inequality becomes an equality.
\label{cpm}
\end{lemma}

\begin{proof}
See~\cite{nielsen:00} page 406.
\end{proof}

Another choice for a distance measure is the fidelity.

\begin{definition}
Let $\rho$, $\sigma\in\mathcal{B}(\mathcal{H})$. The fidelity between $\rho$ and $\sigma$ is defined as
\begin{equation}
F(\rho,\sigma)=\left\|\rho^{1/2}\sigma^{1/2}\right\|_{1}^{2}\ .
\end{equation}
\end{definition}

\begin{lemma}
Let $\rho$, $\sigma\in\mathcal{B}(\mathcal{H})$ and let $\Lambda$ be a CPTP map. Then
\begin{equation}
F(\rho,\sigma)\leq F(\Lambda(\rho),\Lambda(\sigma))\ .
\end{equation}
In addition, if $\Lambda$ is an isometry then the inequality becomes an equality.
\label{uhhh}
\end{lemma}

\begin{proof}
See~\cite{jozsa:94}.
\end{proof}

The trace distance and the fidelity are qualitatively equivalent measures of closeness for quantum states.

\begin{lemma}
Let $\rho$, $\sigma\in\mathcal{B}(\mathcal{H})$. The fidelity is related to the trace norm as follows
\begin{equation}
1-\sqrt{F(\rho,\sigma)}\leq\frac{1}{2}\left\|\sigma-\rho\right\|_{1}\leq\sqrt{1-F(\rho,\sigma)}\ .
\end{equation}
\label{fidtr}
\end{lemma}

\begin{proof}
See~\cite{fuchs:99}.
\end{proof}

\chapter{(Smooth) Min-/Max-Entropy and Collision Entropy}
\label{sminmax}
\lhead{Chapter 3. \emph{(Smooth) Min-/Max-Entropy}}

Smooth min- and max-entropy were introduced in~\cite{renner:05, renner:04, wolf:04} and can be seen as generalizations of the von Neumann entropy.

As we will see smooth min- and max-entropy are the entropy measures that quantify the so called minimal entanglement cost in the problem of quantum state merging (cp. Chapter 4). For a further motivation of the definitions and a more extensive treatment see~\cite{renner:05, renner:04}. 

\section{Min- and Max-Entropy}

In this section we introduce a non-smooth version of min- and max-entropy. It is the basis for the definition of smooth min- and max-entropy in Section 3.2. We first give a definition for the unconditional min- and max-entropy.

\begin{definition}
Let $\rho\in\mathcal{B}(\mathcal{H})$. The min- and max-entropy of $\rho$ are defined by
\begin{align}
& H_{\min}(\rho)=-\log{\lambda_{\max}(\rho)}\\
& H_{\max}(\rho)=\log{\textrm{rank}(\rho)}\ ,
\end{align}
where $\lambda_{\max}(.)$ denotes the maximum eigenvalue of the argument.
\end{definition}

Note that these are special cases of the quantum $\alpha$-Renyi entropy $H_{\alpha}=\frac{1}{1-\alpha}\log{\textrm{tr}(\rho^{\alpha})}$, where $\alpha\geq0$. Namely we can get $H_{\max}$ for $\alpha\rightarrow0$ and $H_{\min}$ for $\alpha\rightarrow\infty$.

\begin{definition}
Let $\rho_{AB}\in\mathcal{B}(\mathcal{H}_{A}\otimes\mathcal{H}_{B})$ and $\sigma_{B}\in\mathcal{B}(\mathcal{H}_{B})$. The conditional min-entropy of $\rho_{AB}$ relative to $\sigma_{B}$ is defined by
\begin{equation}
H_{\min}(\rho_{AB}|\sigma_{B})=-\log{\lambda}\ ,
\end{equation}
where $\lambda$ is the minimum real number such that $\lambda \cdot \id_{A}\otimes \sigma_{B} -\rho_{AB}$ is non-negative. The conditional max-entropy of $\rho_{AB}$ relative to $\sigma_{B}$ is defined by
\begin{equation}
H_{\max}(\rho_{AB}|\sigma_{B})=\log{\textrm{tr}((\id_{A}\otimes\sigma_{B})\rho_{AB}^{0})}\ ,
\end{equation}
where $\rho_{AB}^{0}$ denotes the projector onto the support of $\rho_{AB}$.
\label{min}
\end{definition}

\begin{definition}
Let $\rho_{AB}\in\mathcal{B}(\mathcal{H}_{A}\otimes\mathcal{H}_{B})$. The min- and max-entropy of $\rho_{AB}$ given B are
\begin{align}
& H_{\min}(\rho_{AB}| B)=\sup_{\sigma_{B}}H_{\min}(\rho_{AB}|\sigma_{B})\\
& H_{\max}(\rho_{AB}| B)=\sup_{\sigma_{B}}H_{\max}(\rho_{AB}|\sigma_{B})\ ,
\end{align}
where the suprema range over all $\sigma_{B}\in\mathcal{B}(\mathcal{H}_{B})$.
\end{definition}

\begin{remark}
If $\mathcal{H}_{B}$ is the trivial space $\mathbb{C}$, these conditional versions reduce to the unconditional min- and max-entropy.
\end{remark}

\begin{lemma}
Let $\rho_{AB}\in\mathcal{B}(\mathcal{H}_{A}\otimes\mathcal{H}_{B})$ and let $\sigma_{B}\in\mathcal{B}(\mathcal{H}_{B})$ be invertible. Then
\begin{equation}
\begin{split}
H_{\min}(\rho_{AB}|\sigma_{B})&=-\log{\lambda_{\max}((\id_{A}\otimes\sigma_{B}^{-1/2})\rho_{AB}(\id_{A}\otimes \sigma_{B}^{-1/2}))}\\
&=-\log{\underset{\vartheta_{AB}}{\textrm{max }}\textrm{tr}(\vartheta_{AB}(\id_{A}\otimes \sigma_{B}^{-1/2})\rho_{AB}(\id_{A}\otimes \sigma_{B}^{-1/2}))}\ ,
\end{split}
\end{equation}
where the maximization ranges over all $\vartheta_{AB}\in\mathcal{B}(\mathcal{H}_{A}\otimes\mathcal{H}_{B})$.
\label{mindiff}
\end{lemma}

\begin{proof}
The first equality is Lemma~\ref{ja} with $\sigma =\id_{A}\otimes\sigma_{B}$ and $\rho=\rho_{AB}$. The second one is an immediate consequence of the first.
\end{proof}

\begin{remark}
Even if $\sigma_{B}$ is not invertible, we can sometimes use a version of Lemma~\ref{mindiff} as well. Consider $\sigma_{B}\in\mathcal{B}(\mathcal{H}_{B})$, $\rho_{AB}\in\mathcal{B}(\mathcal{H}_{A}\otimes\mathcal{H}_{B})$ with $\textrm{supp}\left\{\textrm{tr}_{A}(\rho_{AB})\right\}\subseteq\textrm{supp}\left\{\sigma_{B}\right\}$ and denote the projector onto the support of $\sigma_{B}$ by $\sigma_{B}^{0}$. To determine $H_{\min}(\rho_{AB}|\sigma_{B})$ we can then read the equation
\begin{equation}
\lambda\cdot\id_{A}\otimes\sigma_{B}\geq\rho_{AB}
\end{equation}
only on the support of $\id_{A}\otimes\sigma_{B}$ because $\rho_{AB}=(\id_{A}\otimes\sigma_{B}^{0})\rho_{AB}(\id_{A}\otimes\sigma_{B}^{0})^{\dagger}$. But on $\textrm{supp}\left\{\sigma_{B}\right\}$, $\sigma_{B}$ has an inverse and we can use this inverse to calculate the min-entropy with Lemma~\ref{mindiff}. So whenever we want to calculate $H_{\min}(\rho_{AB}|\sigma_{B})$ for a $\sigma_{B}$ not invertible but with $\textrm{supp}\left\{\textrm{tr}_{A}(\rho_{AB})\right\}\subseteq\textrm{supp}\left\{\sigma_{B}\right\}$, we denote by $\sigma_{B}^{-1}$ the inverse of $\sigma_{B}$ on $\textrm{supp}\left\{\sigma_{B}\right\}$ and call it generalized inverse of $\sigma_{B}$. We are then allowed to use Lemma~\ref{mindiff}. We especially do this for $\sigma_{B}$ equal to $\rho_{B}$.
\label{achja}
\end{remark}

Min- and max-entropy have many interesting properties. For a more detailed discussion see~\cite{renner:05,renner:04}.

\begin{lemma}[Addidivity]
Let $\rho_{AB}\in\mathcal{B}(\mathcal{H}_{A}\otimes\mathcal{H}_{B})$, $\sigma_{B}\in\mathcal{B}(\mathcal{H}_{B})$ and $\rho_{A'B'}\in\mathcal{B}(\mathcal{H}_{A'}\otimes\mathcal{H}_{B'})$, $\sigma_{B'}\in\mathcal{B}(\mathcal{H}_{B'})$. Then
\begin{align}
& H_{\min}(\rho_{AB}\otimes\rho_{A'B'}|\sigma_{B}\otimes\sigma_{B'})=H_{\min}(\rho_{AB}|\sigma_{B})+H_{\min}(\rho_{A'B'}|\sigma_{B'})\\
& H_{\max}(\rho_{AB}\otimes\rho_{A'B'}|\sigma_{B}\otimes\sigma_{B'})=H_{\max}(\rho_{AB}|\sigma_{B})+H_{\max}(\rho_{A'B'}|\sigma_{B'})\ .
\end{align}
\label{add}
\end{lemma}

\begin{proof}
Clear from Definition~\ref{min}.
\end{proof}

\begin{lemma}[Strong Subadditivity]
Let $\rho_{ABR}\in\mathcal{B}(\mathcal{H}_{A}\otimes\mathcal{H}_{B}\otimes\mathcal{H}_{R})$ and $\sigma_{BR}\in\mathcal{B}(\mathcal{H}_{B}\otimes\mathcal{H}_{R})$. Then
\begin{align}
& H_{\min}(\rho_{ABR}|\sigma_{BR})\leq H_{\min}(\rho_{AB}|\sigma_{B})\\
& H_{\max}(\rho_{ABR}|\sigma_{BR})\leq H_{\max}(\rho_{AB}|\sigma_{B})\ .
\end{align}
\label{strong}
\end{lemma}

\begin{proof}
See Lemma 3.1.7 in~\cite{renner:05}.
\end{proof}

\begin{lemma}
Let $\rho_{ABR}\in\mathcal{B}(\mathcal{H}_{A}\otimes\mathcal{H}_{B}\otimes\mathcal{H}_{R})$ and denote the dimension of $\mathcal{H}_{B}$ by $d_{B}$. Then
\begin{equation}
H_{\min}(\rho_{ABR}|R)\leq H_{\min}(\rho_{AR}|R)+\log d_{B}\ .
\end{equation}
\label{neusa}
\end{lemma}

\begin{proof}
Let $H_{\min}(\rho_{ABR}|R)=H_{\min}(\rho_{ABR}|\overline{\sigma}_{R})=-\log\lambda$, i.e.~$\lambda$ is minimal such that $\lambda\cdot\id_{AB}\otimes\overline{\sigma}_{R}\geq\rho_{ABR}$. By taking the partial trace over $B$ we get $\lambda\cdot d_{B}\cdot\id_{A}\otimes\overline{\sigma}_{R}\geq\rho_{AR}$. Furthermore we have $H_{\min}(\rho_{AR}|R)\geq H_{\min}(\rho_{AR}|\overline{\sigma}_{R})=-\log\mu$, where $\mu$ is minimal such that $\mu\cdot\id_{A}\otimes\overline{\sigma}_{R}\geq\rho_{AR}$. Hence $\lambda\cdot d_{B}\geq\mu$ and therefore
\begin{equation}
H_{\min}(\rho_{ABR}|R)\leq H_{\min}(\rho_{AR}|\overline{\sigma}_{R})+\log d_{B}\leq H_{\min}(\rho_{AR}|R)+\log d_{B}\ .
\end{equation}
\end{proof}

If we condition on the reduced density matrix, we can get a very simple formula for the min-entropy 
of pure states.

\begin{lemma}
Let $\rho_{AB}\in\mathcal{B}(\mathcal{H}_{A}\otimes\mathcal{H}_{B})$ with $\rho_{AB}=\ket{\psi}\bra{\psi}_{AB}$. Then
\begin{equation}
H_{\min}(\rho_{AB}|\rho_{B})=-\log{r}
\end{equation}
where $r$ is the Schmidt-rank of $\ket{\psi_{AB}}$ (cp.~Lemma~\ref{schmidt}).
\label{krass}
\end{lemma}

\begin{proof}
Write $H_{\min}(\rho_{AB}|\rho_{B})=-\log{\lambda}$ and due to Lemma~\ref{mindiff} it remains to prove
\begin{equation}
r=\lambda_{\max}((\id_{A}\otimes\rho_{B}^{-1/2})\rho_{AB}(\id_{A}\otimes\rho_{B}^{-1/2}))\ .
\label{ghgh}
\end{equation}
Now use a Schmidt-decomposition of $\ket{\psi}_{AB}$ with Schmidt-coefficients $\lambda_{i}$ and calculate the right-hand side of (\ref{ghgh})
\begin{equation}
\begin{split}
\lambda_{\max}&((\sum_{m}\ket{m}\bra{m}_{A}\otimes\sum_{\underset{\lambda_{n}\neq0}{n}}\lambda_{n}^{-1/2}\ket{n}\bra{n}_{B})(\sum_{ij}\sqrt{\lambda_{i}\lambda_{j}}\ket{ii}\bra{jj}_{AB})\\
&(\sum_{k}\ket{k}\bra{k}_{A}\otimes\sum_{\underset{\lambda_{l}\neq0}{l}}\lambda_{l}^{-1/2}\ket{l}\bra{l}_{B}))\\
&=\lambda_{\max}(\sum_{\underset{\lambda_{i}\lambda_{j}\neq 0}{ij}}\ket{ii}\bra{jj}_{AB}).
\end{split}
\end{equation}
The only eigenvector of
\begin{equation}
\sum_{\underset{\lambda_{i}\lambda_{j}\neq0}{ij}}\ket{ii}\bra{jj}_{AB}
\end{equation}
with non-zero eigenvalue is
\begin{equation}
\ket{\xi}=\sum_{\underset{\lambda_{k}\neq0}{k}}\ket{kk}_{AB}\ .
\end{equation}
The corresponding eigenvalue $\lambda_{\xi}=\lambda_{\max}$ can be determined by
\begin{equation}
\lambda_{\xi}(\sum_{\underset{\lambda_{k}\neq0}{k}}\ket{kk}_{AB})=(\sum_{\underset{\lambda_{i}\lambda_{j}\neq 0}{ij}}\ket{ii}\bra{jj}_{AB})(\sum_{\underset{\lambda_{k}\neq0}{k}}\ket{kk}_{AB})\ .
\end{equation}
This implies $\lambda_{\xi}=r$.
\end{proof}

The min- and max-entropy are dual to each other in the following sense.

\begin{proposition}
Let $\rho_{ABR}\in\mathcal{B}(\mathcal{H}_{A}\otimes\mathcal{H}_{B}\otimes\mathcal{H}_{R})$ with $\rho_{ABR}=\ket{\psi}\bra{\psi}_{ABR}$. Then
\begin{equation}
H_{\min}(\rho_{AR}|\rho_{R})=-H_{\max}(\rho_{AB}| B)\ .
\end{equation}
\label{minmax}
\end{proposition}

\begin{proof}
Due to Lemma~\ref{mindiff} we can get 
\begin{equation}
2^{-H_{\min}(\rho_{AR}|\rho_{R})}=\lambda_{\mathrm{max}}((\id_{A}\otimes\rho_{R}^{-1/2})\rho_{AR}(\id_{A}\otimes\rho_{R}^{-1/2}))\ .
\end{equation}
Now define $\omega_{ABR}=(\id_{AB}\otimes\rho_{R}^{-1/2})\rho_{ABR}(\id_{AB}\otimes\rho_{R}^{-1/2})$ and note that $\omega_{ABR}$ is pure since $\rho_{ABR}$ is pure. A Schmidt-decomposition of $\omega_{ABR}$ into $AR$, $B$ gives us that
\begin{equation}
\lambda_{\mathrm{max}}((\id_{A}\otimes\rho_{R}^{-1/2})\rho_{AR}(\id_{A}\otimes\rho_{R}^{-1/2}))=\lambda_{\mathrm{max}}(\omega_{AR})=\lambda_{\mathrm{max}}(\omega_{B})\ .
\end{equation}
Using Lemma~\ref{mindiff} we get
\begin{equation}
\lambda_{\mathrm{max}}(\omega_{B})=\max_{\sigma_{B}}\mathrm{tr}(\sigma_{B}\omega_{B})=\max_{\sigma_{B}}\mathrm{tr}((\id_{A}\otimes\sigma_{B})\omega_{AB})\ ,
\label{neuneu}
\end{equation}
where the maximization ranges over all $\sigma_{B}\in\mathcal{B}(\mathcal{H}_{B})$. A Schmidt-decomposition of $\ket{\psi}_{ABR}$ into $AB$, $R$ let's us see that
\begin{equation}
(\id_{AB}\otimes\rho_{R}^{-1/2})\ket{\psi}_{ABR}=\sum_{i}\ket{i}_{AB}\otimes\ket{i}_{R}=:\ket{\Phi}_{ABR}\ .
\end{equation}
Since $\ket{\Phi}_{ABR}$ is a fully entangled state we have that
\begin{equation}
(\id_{AB}\otimes\rho_{R}^{-1/2})\ket{\Phi}_{ABR}=((\rho_{AB}^{-1/2})^{T}\otimes\id_{R})\ket{\Phi}_{ABR}=(\rho_{AB}^{-1/2}\otimes\id_{R})\ket{\Phi}_{ABR}\ .
\end{equation}
This implies
\begin{equation}
(\id_{AB}\otimes\rho_{R}^{-1/2})(\id_{AB}\otimes\rho_{R}^{-1/2})\ket{\psi}_{ABR}=(\rho_{AB}^{-1/2}\otimes\rho_{R}^{-1/2})\ket{\psi}_{ABR}
\end{equation}
and by multiplying this with $(\id_{AB}\otimes\rho_{R}^{1/2})$ from the left we get
\begin{equation}
(\id_{AB}\otimes\rho_{R}^{-1/2})\ket{\psi}_{ABR}=(\rho_{AB}^{-1/2}\otimes\id_{R})\ket{\psi}_{ABR}\ .
\end{equation}
Therefore
\begin{equation}
\begin{split}
\omega_{ABR} & =(\id_{AB}\otimes\rho_{R}^{-1/2})\ket{\psi}\bra{\psi}_{ABR}(\id_{AB}\otimes\rho_{R}^{-1/2})\\
& =(\rho_{AB}^{-1/2}\otimes\id_{R})\ket{\psi}\bra{\psi}_{ABR}(\rho_{AB}^{-1/2}\otimes\id_{R})
\end{split}
\end{equation}
and hence
\begin{equation}
\omega_{AB}=\mathrm{tr}_{R}(\omega_{ABR})=\mathrm{tr}_{R}((\rho_{AB}^{-1/2}\otimes\id_{R})\rho_{ABR}(\rho_{AB}^{-1/2}\otimes\id_{R}))=\rho_{AB}^{-1/2}\rho_{AB}\rho_{AB}^{-1/2}=\rho_{AB}^{0}\ .
\end{equation}
Continuing with equation (\ref{neuneu}) we get
\begin{equation}
\begin{split}
\max_{\sigma_{B}}\mathrm{tr}((\id_{A}\otimes\sigma_{B})\omega_{AB})=\max_{\sigma_{B}}\mathrm{tr}((\id_{A}\otimes\sigma_{B})\rho_{AB}^{0}) & \stackrel{\mathrm{(ii)}}{=}\sup_{\sigma_{B}}\mathrm{tr}((\id_{A}\otimes\sigma_{B})\rho_{AB}^{0})\\
& =2^{H_{\max}(\rho_{AB}|B)}\ ,
\end{split}
\end{equation}
where step (ii) is correct since we assumed that all Hilbert spaces are finite dimensional.
\end{proof}

\section{Smooth Min- and Max-Entropy}

Using the definitions of non-smooth min- and max-entropy we now give the definitions for the smooth version. Again, for more details see~\cite{renner:05, renner:04}.

\begin{definition}
Let $\rho_{AB}\in\mathcal{B}(\mathcal{H}_{A}\otimes\mathcal{H}_{B})$, $\sigma_{B}\in\mathcal{B}(\mathcal{H}_{B})$ and $\epsilon\geq 0$. The $\epsilon$-smooth conditional min-entropy and $\epsilon$-smooth conditional max-entropy of $\rho_{AB}$ relative to $\sigma_{B}$ are defined by
\begin{align}
& H_{\min}^{\epsilon}(\rho_{AB}|\sigma_{B})=\sup_{\overline{\rho}_{AB}}H_{\min}(\overline{\rho}_{AB}|\sigma_{B})\\
& H_{\max}^{\epsilon}(\rho_{AB}|\sigma_{B})=\inf_{\overline{\rho}_{AB}}H_{\max}(\overline{\rho}_{AB}|\sigma_{B})\ ,
\end{align}
where the supremeum and the infimum range over all $\overline{\rho}_{AB}\in\mathcal{B}(\mathcal{H}_{A}\otimes\mathcal{H}_{B})$\\
with $\frac{1}{2}\left\|\overline{\rho}_{AB}-\rho_{AB}\right\|_{1} \leq \epsilon$.
\label{drucker}
\end{definition}

\begin{definition}
Let $\rho_{AB}\in\mathcal{B}(\mathcal{H}_{A}\otimes\mathcal{H}_{B})$ and $\epsilon\geq 0$. The $\epsilon$-smooth conditional min-entropy and $\epsilon$-smooth conditional max-entropy of $\rho_{AB}$ given $B$ are defined by
\begin{align}
& H_{\min}^{\epsilon}(\rho_{AB}| B)=\sup_{\sigma_{B}}H_{\min}^{\epsilon}(\rho_{AB}|\sigma_{B})\\
& H_{\max}^{\epsilon}(\rho_{AB}| B)=\sup_{\sigma_{B}}H_{\max}^{\epsilon}(\rho_{AB}|\sigma_{B})\ ,
\end{align}
where the suprema range over all $\sigma_{B}\in\mathcal{B}(\mathcal{H}_{B})$.
\end{definition}

\begin{remark}
We are allowed to restrict the supremum over $\sigma_{B}$ in the definition of the smooth min-entropy to $\sigma_{B}$'s with $\textrm{supp}\left\{\textrm{tr}_{A}(\rho_{AB})\right\}\subseteq\textrm{supp}\left\{\sigma_{B}\right\}$.
\label{neunew}
\end{remark}

Many properties of the non-smooth min- and max-entropy can be generalized to the smooth case.

\begin{lemma}[Superadditivity]
Let $\rho_{AB}\in\mathcal{B}(\mathcal{H}_{A}\otimes\mathcal{H}_{B})$, $\sigma_{B}\in\mathcal{B}(\mathcal{H}_{B})$, $\rho_{A'B'}\in\mathcal{B}(\mathcal{H}_{A'}\otimes\mathcal{H}_{B'})$, $\sigma_{B'}\in\mathcal{B}(\mathcal{H}_{B'})$ and $\epsilon, \epsilon '\geq 0$. Then
\begin{equation}
H_{\min}^{\epsilon+\epsilon '}(\rho_{AB}\otimes\rho_{A'B'}|\sigma_{B}\otimes\sigma_{B'})\geq H_{\min}^{\epsilon}(\rho_{AB}|\sigma_{B})+H_{\min}^{\epsilon '}(\rho_{A'B'}|\sigma_{B'})\ .
\end{equation}
\end{lemma}

\begin{proof}
See Lemma 3.2.6 in~\cite{renner:05}.
\end{proof}

\begin{lemma}[Strong Subadditivity]
Let $\rho_{ABR}\in\mathcal{B}(\mathcal{H}_{A}\otimes\mathcal{H}_{B}\otimes\mathcal{H}_{R})$, $\sigma_{BR}\in\mathcal{B}(\mathcal{H}_{B}\otimes\mathcal{H}_{R})$ and $\epsilon \geq 0$. Then
\begin{equation}
H_{\min}^{\epsilon}(\rho_{ABR}|\sigma_{BR})\leq H_{\min}^{\epsilon}(\rho_{AB}|\sigma_{B})\ .
\end{equation}
\end{lemma}

\begin{proof}
See Lemma 3.2.7 in~\cite{renner:05}.
\end{proof}

\begin{lemma}
Let $\rho_{ABR}\in\mathcal{B}(\mathcal{H}_{A}\otimes\mathcal{H}_{B}\otimes\mathcal{H}_{R})$, denote the dimension of $\mathcal{H}_{B}$ by $d_{B}$ and let $\epsilon\geq0$. Then
\begin{equation}
H_{\min}^{\epsilon}(\rho_{ABR}|R)\leq H_{\min}^{\epsilon}(\rho_{AR}|R)+\log d_{B}\ .
\end{equation}
\label{neusi}
\end{lemma}

\begin{proof}
Let $H_{\min}^{\epsilon}(\rho_{ABR}|R)=H_{\min}(\sigma_{ABR}|R)$ and hence $\|\rho_{ABR}-\sigma_{ABR}\|_{1}\leq2\epsilon$. Lemma~\ref{neusa} gives us
\begin{equation}
H_{\min}(\sigma_{ABR}|R)\leq H_{\min}(\sigma_{AR}|R)+\log d_{B}\stackrel{\mathrm{(i)}}{\leq}H_{\min}^{\epsilon}(\rho_{AR}|R)+\log d_{B}\ ,
\end{equation}
where step (i) is correct since the trace distance does not increase under CPTP maps (Lemma~\ref{cpm}) and hence $\|\rho_{AR}-\sigma_{AR}\|_{1}\leq2\epsilon$.
\end{proof}

Smooth conditional min- and max-entropy of product states are asymptotically equal to the conditional von Neumann entropy. This statement is made precise in Theorem 3.3.6 in~\cite{renner:05}.

\begin{remark}
Since all Hilbert spaced are assumed to be finite dimensional, all suprema and infima can be replaced by maxima and minima resp.
\end{remark}

\section{Collision Entropy}

For technical reasons we will also need the collision entropy. It is a generalization of the classical condition collision entropy to quantum states.

\begin{definition}
Let $\rho_{AB}\in\mathcal{B}(\mathcal{H}_{A}\otimes\mathcal{H}_{B})$ and $\sigma_{B}\in\mathcal{B}(\mathcal{H}_{B})$. The conditional collision entropy of $\rho_{AB}$ relative to $\sigma_{B}$ is defined by
\begin{equation}
H_{2}(\rho_{AB}|\sigma_{B})=-\log{\textrm{tr}(((\id_{A}\otimes\sigma_{B}^{-1/4})\rho_{AB}(\id_{A}\otimes\sigma_{B}^{-1/4}))^{2})}\ ,
\end{equation}
where $\sigma_{B}^{-1}$ denotes the generalized inverse of $\sigma_{B}$.
\end{definition}

\begin{lemma}
Let $\rho_{AB}\in\mathcal{B}(\mathcal{H}_{A}\otimes\mathcal{H}_{B})$ and $\sigma_{B}\in\mathcal{B}(\mathcal{H}_{B})$ with $\textrm{supp}\left\{\textrm{tr}_{A}(\rho_{AB})\right\}\subseteq\textrm{supp}\left\{\sigma_{B}\right\}$. Then
\begin{equation}
H_{\min}(\rho_{AB}|\sigma_{B})\leq H_{2}(\rho_{AB}|\sigma_{B})\ .
\end{equation}
\label{kopf}
\end{lemma}

\begin{proof}
With Lemma~\ref{mindiff} the assertion becomes equivalent to the trivial statement
\begin{equation}
\underset{\vartheta_{AB}}{\textrm{max }}\textrm{tr}(\vartheta_{AB}(\id_{A}\otimes\sigma_{B}^{-1/2})\rho_{AB}(\id_{A}\otimes \sigma_{B}^{-1/2}))\geq \textrm{tr}(\rho_{AB}(\id_{A}\otimes\sigma_{B}^{-1/2})\rho_{AB}(\id_{A}\otimes\sigma_{B}^{-1/2}))\ ,
\end{equation}
where $\vartheta_{AB}\in \mathcal{B}(\mathcal{H}_{A}\otimes\mathcal{H}_{B})$.
\end{proof}

\chapter{Quantum State Merging}
\label{merg}
\lhead{Chapter 4. \emph{Quantum State Merging}}

We consider a quantum information source that emits a sequence of pure states $\ket{\psi^{1}}_{AB}$, $\ket{\psi^{2}}_{AB}$, $\ldots$ with average density matrix $\rho_{AB}$ and assume that the statistics of the source are known to Alice and Bob but not the actual sequence. We allow classical communication for free and ask how much quantum communication is needed to transfer any sequence of pure states that realizes $\rho_{AB}$ to Bob. Since we allow classical communication for free we can replace quantum communication by entanglement due to teleportation~\cite{bennett:93}. This appears to be a more comprehensible way of thinking of the quantum communication.

Moreover there is an equivalent but much more elegant way to think of this problem. We can imagine that $\rho_{AB}$ is part of a larger pure state $\ket{\psi}_{ABR}$ that also lives on a reference system $R$. In this picture faithful state transfer means that Alice can transfer her part of $\ket{\psi}_{ABR}$ to Bob's side and at the same time let the $R$-part of $\ket{\psi}_{ABR}$ unchanged. This motivates the following definition of $\epsilon$-error quantum state merging.

\begin{definition}[Quantum State Merging]
Consider $\rho_{ABR}\in\mathcal{B}(\mathcal{H}_{A}\otimes\mathcal{H}_{B}\otimes\mathcal{H}_{R})$ with $\rho_{ABR}=\ket{\psi}\bra{\psi}_{ABR}$ shared between two parties $A,B$ and a reference $R$. Let $A_{0}$ and $A_{1}$ be further registers at $A$ and $B_{0}$ and $B_{1}$ be further registers at $B$. Furthermore let $B'$ be an ancilla at $B$ of the same size as $A$. A process $\mathcal{M}:AA_{0}\otimes BB_{0}\rightarrow A_{1}\otimes B_{1}B'B$ is called state merging of $\ket{\psi}_{ABR}$ with error $\epsilon\geq0$, if it is a local operation an classical communication process (LOCC), with $\rho_{A_{1}B_{1}B'BR}=(\mathcal{M}\otimes \id_{R})(\ket{\Phi_{K}}\bra{\Phi_{K}}_{A_{0}B_{0}}\otimes\ket{\psi}\bra{\psi}_{ABR})$,
\begin{equation}
\left\|\rho_{A_{1}B_{1}B'BR}-\ket{\Phi_{L}}\bra{\Phi_{L}}_{A_{1}B_{1}}\otimes\ket{\psi}\bra{\psi}_{BB'R}\right\|_{1}\leq\epsilon\ ,
\label{burger}
\end{equation}
with maximally entangled states $\ket{\Phi_{K}},\ket{\Phi_{L}}$ on $A_{0}B_{0},A_{1}B_{1}$ of Schmidt-rank $K$ and $L$, resp. and with $\ket{\psi}_{BB'R}=(\id_{A\rightarrow B'}\otimes\id_{BR})\ket{\psi}_{ABR}$. The number $\log{K}-\log{L}$ is called entanglement cost of the protocol.
\label{cam}
\end{definition}

Our goal is to quantify the minimal entanglement cost for a given $\ket{\psi}_{ABR}$ and $\epsilon$ (or vice versa the minimal $\epsilon$ for given entanglement cost).

\begin{remark}
The term quantum state merging was defined by Horodecki et al.~\cite{horodecki:07} in the same way as we do it here (except that they use the trace distance instead of the fidelity in (\ref{burger})). But they only consider the case of many copies of the same state, $\ket{\psi}_{ABR}=\ket{\varphi^{\otimes n}}_{ABR}$, and analyze what happens for $n\rightarrow\infty$. Because we want to focus on the more general case of an arbitrary $n$ (in particular $n=1$), we henceforth talk about single-shot state merging.
\end{remark}

In Section 4.1 we give a single-shot state merging protocol that achieves $\epsilon$-error merging for a certain entanglement cost. In Section 4.2 we give a general bound for the entanglement cost that shows the optimality of this protocol. Hence we will be able to quantify the minimal entanglement cost. The proofs in this chapter rely on ideas of~\cite{winter:05}.

\section{Single-shot state merging protocol}

Let us first think of a condition that is sufficient to obtain zero error state merging. It is based on a measurement performed on Alice's side, that takes the original state $\ket{\psi}_{ABR}$ to another pure state such that the state on $R$ is unchanged and the state on Alice's side is in product form with the reference's state. Since all purifications are equal up to local unitaries, we can find a local unitary on Bob's side that transforms the state on Bob's side into $\rho_{AB}$.

A more detailed description looks as follows. At the beginning the state is $\ket{\psi}_{ABR}\otimes\ket{\Phi_{K}}_{A_{0}B_{0}}$ and in the end we want it to be $\ket{\psi}_{BB'R}\otimes\ket{\Phi_{L}}_{A_{1}B_{1}}$. We consider a measurement on $AA_{0}$ with operators $P_{j}$ that map $AA_{0}$ to $A_{1}$ and denote the measurement outcomes on $A_{1}BR$ by
\begin{align}
\rho_{A_{1}BR}^{j} & =\ket{\psi^{j}}\bra{\psi^{j}}_{A_{1}BR}\\
& =\textrm{tr}_{B_{0}}(\frac{1}{p_{j}}(P_{j}\otimes \id_{BB_{0}R})(\ket{\psi}\bra{\psi}_{ABR}\otimes\ket{\Phi_{K}}\bra{\Phi_{K}}_{A_{0}B_{0}})(P_{j}\otimes\  \id_{BB_{0}R})^{\dagger})\ ,
\end{align}
where each outcome occurs with probability $p_{j}=\bra{\psi}\otimes\bra{\Phi_{K}}(P_{j}^{\dagger}P_{j}\otimes \id_{BB_{0}R})\ket{\Phi_{K}}\otimes\ket{\psi}$. Now suppose that we have 
\begin{equation}
\rho_{A_{1}R}^{j}=\tau_{A_{1}}\otimes\rho_{R}
\label{ar}
\end{equation}
for each $j$, where $\rho_{R}$ is the reduced density matrix on $R$ of the original state $\ket{\psi}_{ABR}\otimes\ket{\Phi_{K}}_{A_{0}B_{0}}$ and $\tau_{A_{1}}$ is the maximally mixed state of dimension $L$ on $A_{1}$. Then $\ket{\psi^{j}}_{A_{1}BR}$ and $\ket{\Phi_{L}}_{A_{1}B_{1}}\otimes\ket{\psi}_{BB'R}$ are both purifications of $\tau_{A_{1}}\otimes\rho_{R}$. Hence they are related by a local isometry on Bob's side (Uhlmann's theorem~\cite{uhlmann:76, jozsa:94}). I.e.~if we had (\ref{ar}) for $\ket{\psi}_{ABR}$ and $K$, $L$, we could achieve zero error state merging of $\ket{\psi}_{ABR}$ for an entanglement cost of $\log{K}-\log{L}$. For general $\epsilon$-error state merging we can get the following condition.

\begin{proposition}[Merging condition]
Let $\ket{\psi}_{ABR}\otimes\ket{\phi_{K}}_{A_{0}B_{0}}$ be a pure state with $\ket{\phi_{K}}_{A_{0}B_{0}}$ maximally entangled of Schmidt-rank $K$. Consider a measurement on Alice's side with outcomes $j$ which occur with probability $p_{j}$. Denote the state on $A_{1}BR$ after the measurement result $j$ was obtained by $\rho_{A_{1}BR}^{j}=\ket{\psi^{j}}\bra{\psi^{j}}_{A_{1}BR}$ and let $\tau_{A_{1}}$ be the maximally mixed state of dimension $L$ on $A_{1}$. If
\begin{equation}
\sum_{j}p_{j}\|\rho_{A_{1}R}^{j}-\tau_{A_{1}}\otimes\rho_{R}\|_{1}\leq\epsilon\ ,
\label{lenovo}
\end{equation}
where $\rho_{R}$ is the reduced density matrix of $\ket{\psi}_{ABR}$ on $R$, then there exists a $2\sqrt{\epsilon}$-error state merging protocol for $\ket{\psi}_{ABR}$.
\label{lenovo2}
\end{proposition}

\begin{proof}
The line of reasoning is analogue to the zero error case. It follows from Lemma~\ref{fidtr} that
\begin{equation}
\sum_{j}p_{j}\sqrt{F(\rho_{A_{1}R}^{j},\tau_{A_{1}}\otimes\rho_{R})}\geq1-\epsilon/2\ .
\end{equation}
Furthermore
\begin{equation}
\sum_{j}p_{j}F(\rho_{A_{1}R}^{j},\tau_{A_{1}}\otimes\rho_{R})\geq(\sum_{j}p_{j}\sqrt{F(\rho_{A_{1}R}^{j},\tau_{A_{1}}\otimes\rho_{R})})^{2}\geq(1-\epsilon/2)^{2}\geq1-\epsilon\ .
\end{equation}
By Uhlmann's theorem~\cite{uhlmann:76, jozsa:94} there exists isometries $U_{j}$ on Bob's side such that
\begin{equation}
F(\rho_{A_{1}R}^{j},\tau_{A_{1}}\otimes\rho_{R})=F((\id_{A_{1}R}\otimes U_{j})\ket{\psi^{j}}_{A_{1}BR},\ket{\Phi_{L}}_{A_{1}B_{1}}\otimes\ket{\psi}_{BB'R})
\label{achnei}
\end{equation}
and therefore
\begin{equation}
\sum_{j}p_{j}F((\id_{A_{1}R}\otimes U_{j})\ket{\psi^{j}}_{A_{1}BR},\ket{\Phi_{L}}_{A_{1}B_{1}}\otimes\ket{\psi}_{BB'R})\geq1-\epsilon\ .
\end{equation}
Since
\begin{equation}
\begin{split}
&F(\sum_{j}p_{j}(\id_{A_{1}R}\otimes U_{j})\ket{\psi^{j}}\bra{\psi^{j}}_{A_{1}BR}(\id_{A_{1}R}\otimes U_{j})^{\dagger},\ket{\Phi_{L}}_{A_{1}B_{1}}\otimes\ket{\psi}_{BB'R})\\
&=\sum_{j}p_{j}F((\id_{A_{1}R}\otimes U_{j})\ket{\psi^{j}}_{A_{1}BR},\ket{\Phi_{L}}_{A_{1}B_{1}}\otimes\ket{\psi}_{BB'R})\ ,
\end{split}
\end{equation}
it follows that
\begin{equation}
F(\sum_{j}p_{j}(\id_{A_{1}R}\otimes U_{j})\ket{\psi^{j}}\bra{\psi^{j}}_{A_{1}BR}(\id_{A_{1}R}\otimes U_{j})^{\dagger},\ket{\Phi_{L}}_{A_{1}B_{1}}\otimes\ket{\psi}_{BB'R})\geq1-\epsilon\ .
\end{equation}
Finally we can use Lemma~\ref{fidtr} again to rewrite this in terms of the trace distance
\begin{equation}
\|\sum_{j}p_{j}(\id_{A_{1}R}\otimes U_{j})\ket{\psi^{j}}\bra{\psi^{j}}_{A_{1}BR}(\id_{A_{1}R}\otimes U_{j})^{\dagger}-\ket{\Phi_{L}}_{A_{1}B_{1}}\otimes\ket{\psi}_{BB'R}\|_{1}\leq2\sqrt{\epsilon}\ .
\end{equation}
\end{proof}

\begin{remark}
Note that the condition (\ref{lenovo}) must be met for any state merging protocol.
\end{remark}

But how do we realize condition (\ref{lenovo})? The crucial technical result that we will use is the following Lemma about Haar distributed projectors.

\begin{lemma}
Let $\rho_{AR}\in\mathcal{B}(\mathcal{H}_{A}\otimes\mathcal{H}_{R})$, P be a projector from $A$ to $A_{1}$, $U$ be a unitary on $A$, $\omega_{A_{1}R}^{U}=(PU\otimes \id_{R})\rho_{AR}(PU\otimes \id_{R})^{\dagger}$ and $\sigma_{R}\in\mathcal{B}(\mathcal{H}_{R})$. If $U$ is a Haar distributed on $A$ then
\begin{equation}
\left\langle \left\| \frac{d_{A}}{L}\omega_{A_{1}R}^{U}-\tau_{A_{1}}\otimes \rho_{R}\right\|_{1}\right\rangle_{U}\leq 2^{-1/2(H_{2}(\rho_{AR}|\sigma_{R})-\log{L})}\ ,
\end{equation}
where $\tau_{A_{1}}$ is the maximally mixed state of dimension $L$ on $A_{1}$, $d_{A}$ denotes the dimension of $\mathcal{H}_{A}$ and $\left\langle . \right\rangle_{U}$ denotes the average over unitaries $U$.
\label{3}
\end{lemma}

\begin{proof}
Note that it is sufficient to show
\begin{equation}
\left\langle \left\| (\id_{A_{1}}\otimes\sigma_{R}^{-1/4})(\frac{d_{A}}{L}\omega_{A_{1}R}^{U}-\tau_{A_{1}}\otimes \rho_{R})(\id_{A_{1}}\otimes\sigma_{R}^{-1/4})\right\|_{2}^{2}\right\rangle_{U}\leq 2^{-H_{2}(\rho_{AR}|\sigma_{R})}\ .
\label{11}
\end{equation}
The assertion then follows from Lemma~\ref{5.1.3} and Jensen's inequality. To see this put $\sigma=\id_{A_{1}}\otimes\sigma_{R}$ and $S=\frac{d_{A}}{L}\omega_{A_{1}R}^{U}-\tau_{A_{1}}\otimes\rho_{R}$ in Lemma~\ref{5.1.3} and observe that $\textrm{tr}(\id_{A_{1}}\otimes\sigma_{R})=L$.
Define
\begin{equation}
\tilde{\rho}_{AR}=(\id_{A_{1}}\otimes\sigma_{R}^{-1/4})\rho_{AR}(\id_{A_{1}}\otimes\sigma_{R}^{-1/4})
\end{equation}
\begin{equation}
\tilde{\omega}_{A_{1}R}^{U}=(PU\otimes \id_{R})\tilde{\rho}_{AR}(PU\otimes \id_{R})^{\dagger}\ .
\end{equation}
If we insert the definition of $H_{2}(\rho_{AR}|\sigma_{R})$ we can rewrite (\ref{11}) to
\begin{equation}
\left\langle \left\| \frac{d_{A}}{L}\tilde{\omega}_{A_{1}R}^{U}-\tau_{A_{1}}\otimes \tilde{\rho_{R}}\right\|_{2}^{2}\right\rangle_{U}\leq\textrm{tr}(\tilde{\rho}_{AR}^{2})\ .
\label{12}
\end{equation}
It thus remains to show that (\ref{12}) holds. Now note that $\left\langle \tilde{\omega}_{A_{1}R}^{U}\right\rangle_{U}=\frac{L}{d_{A}}\tau_{A_{1}}\otimes\tilde{\rho_{R}}$. Hence the left-hand side of (\ref{12}) has the form of a variance and can be rewritten to
\begin{equation}
\begin{split}
\left\langle \left\| \frac{d_{A}}{L}\tilde{\omega}_{A_{1}R}^{U}-\tau_{A_{1}}\otimes \tilde{\rho_{R}}\right\|_{2}^{2}\right\rangle_{U} & =\frac{d_{A}^{2}}{L^{2}}\left\langle \left\|\tilde{\omega}_{A_{1}R}^{U}-\left\langle \tilde{\omega}_{A_{1}R}^{U}\right\rangle_{U}\right\|_{2}^{2}\right\rangle_{U}\\
&=\frac{d_{A}^{2}}{L^{2}}(\left\langle \textrm{tr}((\tilde{\omega}_{A_{1}R}^{U})^{2})\right\rangle_{U}-\textrm{tr}(\left\langle \tilde{\omega}_{A_{1}R}^{U}\right\rangle_{U}^{2}))\\
&=\frac{d_{A}^{2}}{L^{2}}\left\langle \left\|\tilde{\omega}_{A_{1}R}^{U}\right\|_{2}^{2}\right\rangle_{U}-\frac{1}{L}\textrm{tr}(\tilde{\rho}_{R}^{2})\ .
\label{13}
\end{split}
\end{equation}
To evaluate the first term on the right-hand side of (\ref{13}) we rewrite it in terms of the swap operator $F$ (as in (23) in~\cite{horodecki:07}). We make use of $F_{A_{1}A_{1}}=(P\otimes P)F_{A_{1}A_{1}}(P\otimes P)$ and get
\begin{equation}
\begin{split}
\left\langle \left\| \tilde{\omega}_{A_{1}R}^{U}\right\|_{2}^{2} \right\rangle_{U} &= \left\langle \textrm{tr}((\tilde{\omega}_{A_{1}R}^{U}\otimes\tilde{\omega}_{A_{1}R}^{U})(F_{A_{1}A_{1}}\otimes F_{RR}))\right\rangle_{U}\\
&=\left\langle \textrm{tr}((PU\otimes PU\otimes \id_{RR})(\tilde{\rho}_{AR}\otimes\tilde{\rho}_{AR})(PU\otimes PU \otimes \id_{RR})^{\dagger}(F_{A_{1}A_{1}}\otimes F_{RR}))\right\rangle_{U}\\
&=\textrm{tr}((\tilde{\rho}_{AR}\otimes\tilde{\rho}_{AR})\left\langle (U\otimes U\otimes \id_{RR})^{\dagger}(F_{A_{1}A_{1}}\otimes F_{RR})(U\otimes U\otimes \id_{RR})\right\rangle_{U})\\
&=\textrm{tr}((\tilde{\rho}_{AR}\otimes\tilde{\rho}_{AR})\left\langle (U\otimes U)^{\dagger}F_{A_{1}A_{1}}(U\otimes U)\right\rangle_{U}\otimes F_{RR})\ .
\label{14}
\end{split}
\end{equation}
In Appendix B of~\cite{horodecki:07} it is shown that
\begin{equation}
\left\langle (U\otimes U)^{\dagger}F_{A_{1}A_{1}}(U\otimes U)^{\dagger}\right\rangle_{U}= \frac{L}{d_{A}}\frac{d_{A}-L}{d_{A}^{2}-1}\id_{AA}+\frac{L}{d_{A}}\frac{Ld_{A}-1}{d_{A}^{2}-1}F_{AA}\ .
\end{equation}
We can insert this into equation (\ref{14}) and get
\begin{equation}
\left\langle \left\| \tilde{\omega}_{A_{1}R}^{U}\right\|_{2}^{2} \right\rangle_{U}= \frac{L}{d_{A}}\frac{d_{A}-L}{d_{A}^{2}-1}\textrm{tr}(\tilde{\rho}_{R}^{2})+\frac{L}{d_{A}}\frac{Ld_{A}-1}{d_{A}^{2}-1}\textrm{tr}(\tilde{\rho}_{AR}^{2})
\end{equation}
as well as
\begin{equation}
\frac{d_{A}^{2}}{L^{2}}\left\langle \left\| \tilde{\omega}_{A_{1}R}^{U}\right\|_{2}^{2} \right\rangle_{U}\leq\frac{1}{L}\textrm{tr}(\tilde{\rho}_{R}^{2})+\textrm{tr}(\tilde{\rho}_{AR}^{2})\ .
\end{equation}
Inserting this into (\ref{13}) implies (\ref{12}) and therefore concludes the proof.
\end{proof}

Lemma~\ref{kopf} gives us that Lemma~\ref{3} also holds for $H_{\min}(\rho_{AR}|\sigma_{R})$ instead of\\ $H_{2}(\rho_{AR}|\sigma_{R})$. We can get the following proposition about the feasibility of single-shot quantum state merging.

\begin{lemma}
Let $\rho_{ABR}\in\mathcal{B}(\mathcal{H}_{A}\otimes\mathcal{H}_{B}\otimes\mathcal{H}_{R})$ with $\rho_{ABR}=\ket{\psi}\bra{\psi}_{ABR}$ and $\epsilon>0$. Then there exists a $2\sqrt{2\epsilon}$-error state merging protocol of $\ket{\psi}_{ABR}$ for an entanglement cost $\log{K}-\log{L}=-H_{\min}(\rho_{AR}|R)+2\log{(1/\epsilon)}$.\footnote{Since we need $K,L\in\mathbb{N}$, we can not choose $\log{K}-\log{L}$ exactly equal to $-H_{\min}(\rho_{AR}|R)+2\log{(1/\epsilon)}$ in general. Rather, we need to choose $K,L$ such that $\log{K}-\log{L}$ is minimal but still greater or equal then $-H_{\min}(\rho_{AR}|R)+2\log{(1/\epsilon)}$.}
\label{opop}
\end{lemma}

\begin{proof}
Choose $K$, $L$ such that $\log{K}-\log{L}=-H_{\min}(\rho_{AR}|R)+2\log{(1/\epsilon)}$ and let $\overline{\sigma}_{R}\in\mathcal{B}(\mathcal{H}_{R})$ be such that $H_{\min}(\rho_{AR}|R)=H_{\min}(\rho_{AR}|\overline{\sigma}_{R})$. The starting state is $\ket{\psi}_{ABR}\otimes\ket{\Phi_{K}}_{A_{0}B_{0}}$. Our goal is to define a random measurement according to the description of Lemma~\ref{3} and we do this as follows. Let $d_{A}$ be the local dimension of $\ket{\psi}_{ABR}\otimes\ket{\Phi_{L}}_{A_{0}B_{0}}$ on the $A,A_{0}$ register. Assume for technical reasons that $d_{A}=N\cdot L$ where $N\in\mathbb{N}$. Then we can pick $N$ fixed orthogonal subspaces of dimension $L$.\footnote{In general $d_{A}=N\cdot L+L'$ where $L<L'$. In this case we choose $N-1$ orthogonal subspaces of dimension $L$ and one of dimension $L'$. The argumentation for the proof remains the same, although some coefficients change.} We denote the projectors onto the subspaces followed by a fixed unitary mapping it to $A_{1}$ by $Q_{j},j=1,...,N$. Thereafter we put $P_{j}=Q_{j}U$ with a Haar distributed random unitary $U$ on $AA_{0}$. Lemma~\ref{3} applied to the state $\rho_{AR}\otimes\tau_{A_{0}}$ (where $\tau_{0}$ is the maximally mixed state of dimension $K$ on $A_{0}$) gives us the estimate
\begin{equation}
\begin{split}
\left\langle \sum_{j=1}^{N}\left\| \frac{d_{A}}{L}\omega_{A_{1}R}^{j}-\tau_{A_{1}}\otimes \rho_{R}\right\|_{1}\right\rangle_{U}& \leq N\cdot2^{-1/2(H_{\min}(\rho_{AR}\otimes\tau_{A_{0}}|\overline{\sigma}_{R})-\log{L})}\\
&=N\cdot2^{-1/2(H_{\min}(\rho_{AR}|\overline{\sigma}_{R})+\log{K}-\log{L})}\ .
\end{split}
\end{equation}
where $\omega_{A_{1}R}^{j}=(P_{j}\otimes \id_{R})\rho_{AR}\otimes\tau_{A_{0}}(P_{j}\otimes \id_{R})^{\dagger}$ and $\tau_{A_{1}}$ is the maximally mixed state of dimension $L$ on $A_{1}$. Since $d_{A}=N\cdot L$ and in the notation of Proposition~\ref{lenovo2} $\omega_{A_{1}R}^{j}=p_{j}\rho_{A_{1}R}^{j}$, this is equivalent to
\begin{equation}
\begin{split}
\left\langle \sum_{j=1}^{N}\left\| p_{j}\rho_{A_{1}R}^{j}-\frac{L}{d_{A}}\tau_{A_{1}}\otimes \rho_{R}\right\|_{1}\right\rangle_{U}& \leq 2^{-1/2(H_{\min}(\rho_{AR}|\overline{\sigma}_{R})+\log{K}-\log{L})}\\
&=2^{-1/2(2\log{(1/\epsilon)})}=\epsilon\ .
\label{ach}
\end{split}
\end{equation}
This implies
\begin{equation}
\left\langle \sum_{j=1}^{N} \left|p_{j}-\frac{L}{d_{A}}\right| \right\rangle_{U}\leq\epsilon
\end{equation}
and we obtain
\begin{equation}
\left\langle\sum_{j=1}^{N}p_{j}\left\|\rho_{A_{1}R}^{j}-\tau_{A_{1}}\otimes\rho_{R}\right\|_{1}\right\rangle_{U}\leq2\epsilon\ .
\end{equation}
Now Proposition~\ref{lenovo2} shows that there exists a $2\sqrt{2\epsilon}$-error state merging protocol.
\end{proof}

\begin{proposition}
Let $\rho_{ABR}\in\mathcal{B}(\mathcal{H}_{A}\otimes\mathcal{H}_{B}\otimes\mathcal{H}_{R})$ with $\rho_{ABR}=\ket{\psi}\bra{\psi}_{ABR}$ and $\epsilon>0$. Then there exists a $8\sqrt{\epsilon}$-error state merging protocol of $\ket{\psi}_{ABR}$ for an entanglement cost $\log{K}-\log{L}=-H_{\min}^{\epsilon}(\rho_{AR}|R)+2\log{(1/\epsilon)}$.
\label{das}
\end{proposition}

\begin{proof}
Choose $K$, $L$ such that $\log{K}-\log{L}=-H_{\min}^{\epsilon}(\rho_{AR}|R)+2\log{(1/\epsilon)}$, let $\overline{\sigma}_{R}\in\mathcal{B}(\mathcal{H}_{R})$ such that $H_{\min}^{\epsilon}(\rho_{AR}|R)=H_{\min}^{\epsilon}(\rho_{AR}|\overline{\sigma}_{R})$ and $\overline{\rho}_{AB}\in\mathcal{B}(\mathcal{H}_{A}\otimes\mathcal{H}_{B})$ such that $H_{\min}^{\epsilon}(\rho_{AR}|\overline{\sigma}_{R})=H_{\min}(\overline{\rho}_{AR}|\overline{\sigma}_{R})$. Now the idea is to use the same argumentation as in Lemma~\ref{opop} but for $\overline{\rho}_{AB}$ instead of $\rho_{AB}$. This gives us the estimate
\begin{equation}
\left\langle\sum_{j=1}^{N}\overline{p}_{j}\left\|\overline{\rho}_{A_{1}R}^{j}-\tau_{A_{1}}\otimes\overline{\rho}_{R}\right\|_{1}\right\rangle_{U}\leq2\epsilon\ .
\end{equation}
Using the triangle inequality and Jensen's inequality we can get
\begin{equation}
\begin{split}
\left\langle\sum_{j=1}^{N}p_{j}\|\rho_{A_{1}R}^{j}-\tau_{A_{1}}\otimes\rho_{R}\|_{1}\right\rangle_{U} & \leq\left\langle\sum_{j=1}^{N}\overline{p}_{j}\|\overline{\rho}_{A_{1}R}^{j}-\tau_{A_{1}}\otimes\overline{\rho}_{R}\|_{1}\right\rangle_{U}\\
& +2\left\langle\sum_{j=1}^{N}|p_{j}-\overline{p}_{j}|\right\rangle_{U}\\
& +\left\langle\sum_{j=1}^{N}\overline{p}_{j}\|\overline{\rho}_{A_{1}R}^{j}-\rho_{A_{1}R}^{j}\|_{1}\right\rangle_{U}\\
& +\|\rho_{R}-\overline{\rho}_{R}\|_{1}\\
&\leq2\epsilon+2\cdot2\epsilon+2\epsilon+2\epsilon=10\epsilon\ .
\end{split}
\end{equation}
Now Proposition~\ref{lenovo2} shows that there exists a $2\sqrt{10\epsilon}\leq8\sqrt{\epsilon}$-error state merging protocol.
\end{proof}

\begin{corollary}
Let $\rho_{ABR}\in\mathcal{B}(\mathcal{H}_{A}\otimes\mathcal{H}_{B}\otimes\mathcal{H}_{R})$ with $\rho_{ABR}=\ket{\psi}\bra{\psi}_{ABR}$ and $\epsilon>0$. Then there exists an $\epsilon$-error state merging protocol of $\ket{\psi}_{ABR}$ for an entanglement cost
\begin{equation}
\log{K}-\log{L}=-H_{\min}^{\epsilon^{2}/64}(\rho_{AR}|R)+4\log(\frac{1}{\epsilon})+12\ .
\end{equation}
\label{letzter}
\end{corollary}

\begin{proof}
Straightforward using Proposition~\ref{das}.
\end{proof}

One can either fix $\epsilon$ and then choose $K$, $L$ accordingly or vice versa. This means that you either want to merge a state $\ket{\psi}_{ABR}$ with some maximal error $\epsilon$ or as accurate as possible for some amount of entanglement available.

\section{General bounds for state merging}

To show that the protocol described in Section 4.1 is tight, we try to find a general bound of the form
\begin{equation}
\log{K}-\log{L}\geq -H_{\min}^{\epsilon}(\rho_{AR}| R)
\end{equation}
for $\epsilon$-error state merging.

In order to obtain such a bound we first analyze the zero error case. Quantum state merging is by definition LOCC on $A$, $B$. So if we look at the part $AR$, quantum state merging only acts on $A$. Hence we try to find an amplitude that is monotone under local operations on $A$ and involves the conditional min-entropy.

\begin{proposition}
Let $\rho_{AR}\in\mathcal{B}(\mathcal{H}_{A}\otimes\mathcal{H}_{R})$, $\sigma_{R}\in\mathcal{B}(\mathcal{H}_{R})$ with $\textrm{supp}\left\{\textrm{tr}_{A}(\rho_{AR})\right\}\subseteq\textrm{supp}\left\{\sigma_{R}\right\}$ and let $\Lambda=(\Lambda_{A}\otimes\id_{R}):AR\rightarrow AR$ be a local operation on $A$ with $\Lambda(\rho_{AR})=\sum_{x}p_{x}\rho_{AR}^{x}$. Then
\begin{equation}
H_{\min}(\rho_{AR}|\sigma_{R})\geq H_{\min}(\rho'_{ARX}|\sigma_{R}\otimes\rho_{X})
\end{equation}
where $\mathcal{H}_{X}$ is an ancilla system with mutually orthogonal basis $\left\{\ket{x}\right\}_{x\in X}$ that corresponds to the measurement outcomes of the local operation $\Lambda$, $\rho'_{ARX}=\sum_{x}p_{x}\rho_{AR}^{x}\otimes\ket{x}\bra{x}$ and $\rho_{X}=\sum_{x}p_{x}\ket{x}\bra{x}$.
\label{sexy}
\end{proposition}

\begin{proof}
We do the proof in three steps. First we show the monotonicity property for unitaries (a), then for projective measurements (b) and finally for general measurements (c).

(a) Write $H_{\min}(\rho_{AR}|\sigma_{R})=-\log{\lambda}$, i.e.~$\lambda$ is minimal such that
\begin{equation}
\rho_{AR}\leq\lambda\cdot\id_{A}\otimes\sigma_{R}\ .
\label{wetzikon}
\end{equation}
Consider a unitary evolution $U_{A}$ on system $A$ and apply the unitary operator $(U_{A}\otimes\id_{R})$ to both sides of (\ref{wetzikon})
\begin{equation}
\begin{split}
(U_{A}\otimes\id_{R})\rho_{AR}(U_{A}\otimes\id_{R})^{\dagger}& \leq\lambda\cdot(U_{A}\otimes\id_{R})\id_{A}\otimes\sigma_{R}(U_{A}\otimes\id_{R})^{\dagger}\\
&=\lambda\cdot\id_{A}\otimes\sigma_{R}\ .
\end{split}
\end{equation}
Now set $H_{\min}((U_{A}\otimes\id_{R})\rho_{AR}(U_{A}\otimes\id_{R})^{\dagger}|\sigma_{R})=-\log{\lambda'}$, i.e.~$\lambda'$ is minimal such that
\begin{equation}
(U_{A}\otimes\id_{R})\rho_{AR}(U_{A}\otimes\id_{R})^{\dagger}\leq\lambda'\cdot\id_{A}\otimes\sigma_{R}\ .
\end{equation}
Hence we have $\lambda=\lambda'$ and therefore
\begin{equation}
H_{\min}(\rho_{AR}|\sigma_{R})=H_{\min}((U_{A}\otimes\id_{R})\rho_{AR}(U_{A}\otimes\id_{R})^{\dagger}|\sigma_{R})\ .
\end{equation}

(b) Consider a projective measurement with projectors $\{P_{A}^{x}\}_{x\in X}$ and let $H_{\min}(\rho_{AR}|\sigma_{R})=-\log{\lambda}$, $H_{\min}(\rho'_{ARX}|\sigma_{R}\otimes\rho_{X})=-\log{\lambda'}$ and $H_{\min}(\rho_{AR}^{x}|\sigma_{R})=-\log{\lambda_{x}}$. We first like to rewrite $\lambda'$ in terms of the $\lambda_{x}$. Because the vectors $\ket{x}$ are mutually orthogonal, the equivalence
\begin{equation}
\mu\cdot\id_{A}\otimes\sigma_{R}\otimes\rho_{X}-\rho'_{ARX}\geq0\Leftrightarrow\forall x:\mu\cdot\id_{A}\otimes\sigma_{R}-\rho_{AR}^{x}\geq0
\label{crack}
\end{equation}
holds for any $\mu\geq 0$. If we take $\mu$ minimal such that (\ref{crack}) holds, we get $\lambda'=\mu=\underset{x}{\textrm{max }}\lambda_{x}$. Thus the assertion becomes equivalent to
\begin{equation}
\lambda\leq\underset{x}{\textrm{max }}\lambda_{x}\ .
\label{zuzu}
\end{equation}
Now let $\rho_{ARE}=\ket{\psi}\bra{\psi}_{ARE}$ be a purification of $\rho_{AR}$ and let $\ket{\psi^{x}}_{ARE}=1/\sqrt{p_{x}}(P_{A}^{x}\otimes\id_{RE})\ket{\psi}_{ARE}$. Note that $\ket{\psi}_{ARE}=\sum_{x}\sqrt{p_{x}}\ket{\psi^{x}}_{ARE}$. Furthermore define $\omega_{ARE}=(\id_{AE}\otimes\sigma_{R}^{-1/2})\ket{\psi}\bra{\psi}_{ARE}(\id_{AE}\otimes\sigma_{R}^{-1/2})$ and $\omega_{ARE}^{x}=(\id_{AE}\otimes\sigma_{R}^{-1/2})\ket{\psi^{x}}\bra{\psi^{x}}_{ARE}(\id_{AE}\otimes\sigma_{R}^{-1/2})$ which are both a pure. Using Lemma~\ref{mindiff} we can get
\begin{equation}
\begin{split}
\lambda & =\lambda_{\max}((\id_{A}\otimes\sigma_{R}^{-1/2})\rho_{AR}(\id_{A}\otimes\sigma_{R}^{-1/2}))=\lambda_{\max}(\omega_{AR})\stackrel{\mathrm{(i)}}{=}\lambda_{\max}(\omega_{E})\\
& =\max_{\sigma_{E}}\mathrm{tr}(\sigma_{E}\omega_{E})=\max_{\sigma_{E}}\mathrm{tr}((\id_{AR}\otimes\sigma_{E})\omega_{ARE})\\
&
=\max_{\sigma_{E}}\mathrm{tr}((\id_{AR}\otimes\sigma_{E})(\id_{AE}\otimes\sigma_{R}^{-1/2})\ket{\psi}\bra{\psi}_{ARE}(\id_{AE}\otimes\sigma_{R}^{-1/2}))\\
& =\max_{\sigma_{E}}\sum_{xx'}\sqrt{p_{x}p_{x'}}\cdot\mathrm{tr}((\id_{AR}\otimes\sigma_{E})(\id_{AE}\otimes\sigma_{R}^{-1/2})\ket{\psi^{x}}\bra{\psi^{x'}}_{ARE}(\id_{AE}\otimes\sigma_{R}^{-1/2}))\\
& \stackrel{\mathrm{(ii)}}{=}\max_{\sigma_{E}}\sum_{x}p_{x}\cdot\mathrm{tr}((\id_{AR}\otimes\sigma_{E})(\id_{AE}\otimes\sigma_{R}^{-1/2})\ket{\psi^{x}}\bra{\psi^{x}}_{ARE}(\id_{AE}\otimes\sigma_{R}^{-1/2}))\\
& =\max_{\sigma_{E}}\sum_{x}p_{x}\cdot\mathrm{tr}((\id_{AR}\otimes\sigma_{E})\omega_{ARE}^{x})\leq\sum_{x}p_{x}\cdot\max_{\sigma_{E}}\mathrm{tr}((\id_{AR}\otimes\sigma_{E})\omega_{ARE}^{x})\\
& =\sum_{x}p_{x}\cdot\max_{\sigma_{E}}\mathrm{tr}(\sigma_{E}\omega_{E}^{x})=\sum_{x}p_{x}\cdot\lambda_{\max}(\omega_{E}^{x})\stackrel{\mathrm{(iii)}}{=}\sum_{x}p_{x}\cdot\lambda_{\max}(\omega_{AR}^{x})\\
& =\sum_{x}p_{x}\cdot\lambda_{\max}((\id_{A}\otimes\sigma_{R}^{-1/2})\rho_{AR}^{x}(\id_{A}\otimes\sigma_{R}^{-1/2}))=\sum_{x}p_{x}\cdot\lambda_{x}\leq\max_{x}\lambda_{x}\ ,
\end{split}
\end{equation}
where the maximization ranges over all $\sigma_{E}\in\mathcal{B}(\mathcal{H}_{E})$. A Schmidt-decomposition of $\omega_{ARE}$ into $AR$, $E$ justifies step (i). To see that step (ii) is correct first note that the $\ket{\psi^{x}}_{ARE}$ are mutually orthogonal. It follows that the $(\id_{AE}\otimes\sigma_{R}^{-1/2})\ket{\psi^{x}}_{ABR}$ are also mutually orthogonal since $(P_{A}^{x}\otimes\id_{RE})$ and $(\id_{AE}\otimes\sigma_{R}^{-1/2})$ commute. Because the operator $(\id_{AR}\otimes\sigma_{E})$ only acts nontrivially on $E$ (ii) holds. Finally a Schmidt-decomposition of $\omega_{ARE}^{x}$ into $E$, $AR$ justifies step (iii).

(c) It is shown in Lemma~\ref{general} that projective measurements together with unitary dynamics are sufficient to implement general measurements if we allow to introduce an extra quantum system (see Lemma~\ref{general} for details). Let $A'$ be this extra system and denote the state on $AA'R$ at the beginning by $\rho_{AR}\otimes\ket{\varphi}\bra{\varphi}_{A'}$. Lemma~\ref{add} gives us that $H_{\min}(\rho_{AR}|\sigma_{R})=H_{\min}(\rho_{AR}\otimes\ket{\varphi}\bra{\varphi}_{A'}|\sigma_{R})$. After applying the projective measurement and the unitary that model the general measurement, we get
\begin{equation}
H_{\min}(\rho_{AR}\otimes\ket{\varphi}\bra{\varphi}_{A'}|\sigma_{R})\geq H_{\min}(\sum_{x}p_{x}\rho_{AR}^{x}\otimes\ket{\varphi^{x}}\bra{\varphi^{x}}_{A'}\otimes\ket{x}\bra{x}|\sum_{x}p_{x}\sigma_{R}\otimes\ket{x}\bra{x})
\label{fuchs}
\end{equation}
because of (a) and (b). Due to an analogue argumentation as at the beginning of step (b) and Lemma~\ref{add} the right-hand side of (\ref{fuchs}) is equal to
\begin{equation}
\underset{x}{\textrm{min }}H_{\min}(\rho_{AR}^{x}\otimes\ket{\varphi^{x}}\bra{\varphi^{x}}_{A'}\otimes\ket{x}\bra{x}|\sigma_{R})=\underset{x}{\textrm{min }}H_{\min}(\rho_{AR}^{x}|\sigma_{R})=H_{\min}(\rho'_{ARX}|\sigma_{R}\otimes\rho_{X})\ .
\end{equation}
This concludes the proof.
\end{proof}

\begin{proposition}
Let $\rho_{ABR}\in\mathcal{B}(\mathcal{H}_{A}\otimes\mathcal{H}_{B}\otimes\mathcal{H}_{R})$ with $\rho_{ABR}=\ket{\psi}\bra{\psi}_{ABR}$. Then it holds for any zero error quantum state merging of $\ket{\psi}_{ABR}$ that
\begin{equation}
\log{K}-\log{L}\geq-H_{\min}(\rho_{AR}|\rho_{R})\ .
\label{cooop}
\end{equation}
\label{spar}
\end{proposition}

\begin{proof}
The initial state is $\ket{\psi}_{ABR}\otimes\ket{\Phi_{K}}_{A_{0}B_{0}}$ and the final state is $\ket{\psi}_{BB'R}\otimes\ket{\Phi_{L}}_{A_{0}B_{0}}$. Proposition~\ref{sexy} applied to the $AR$-part for $\sigma_{R}=\rho_{R}$ gives
\begin{equation}
H_{\min}(\rho_{AR}\otimes\tau_{A_{0}}|\rho_{R})\geq H_{\min}(\sum_{x}p_{x}\tau_{A_{1}}\otimes\rho_{R}\otimes\ket{x}\bra{x}|\sum_{x}p_{x}\rho_{R}\otimes\ket{x}\bra{x})\ ,
\label{migros}
\end{equation}
where the $x$ denote the measurement outcomes of the local operations on $A$ (of any hypothetical state merging protocol). The left-hand side of (\ref{migros}) is equal to $H_{\min}(\rho_{AR}|\rho_{R})+\log{K}$ and the right-hand side of (\ref{migros}) is equal to $\log{L}$. This concludes the proof.
\end{proof}

\begin{remark}
Note that our definition of state merging does not allow Alice to use any additional register on her side (e.g.~a random bit). However the bound (\ref{cooop}) still holds if we allow this.
\begin{proof}
Denote the state on the additional register at the beginning by $\rho'_{A}$. In the picture of state merging we need to think of this as a pure state $\ket{\varphi}_{A'R'}$ that also lives on a reference system $R'$. Hence the state at the beginning is given by $\ket{\psi}_{ABR}\otimes\ket{\Phi_{K}}_{A_{0}B_{0}}\otimes\ket{\varphi}_{A'R'}$. Proposition~\ref{sexy} gives us
\begin{equation}
\log{K}-\log{L}\geq-H_{\min}(\rho_{AR}|\rho_{R})-H_{\min}(\ket{\varphi}\bra{\varphi}_{A'R'}|\rho_{R'})\ .
\label{denner}
\end{equation}
But Lemma~\ref{krass} tells us that $H_{\min}(\ket{\varphi}\bra{\varphi}_{A'R'}|\rho_{R'})=-\log{r}$, where $r$ is the Schmidt-rank of $\ket{\varphi}_{A'R'}$. Hence the right-hand side of (\ref{denner}) is always greater or equal than $-H_{\min}(\rho_{AR}|\rho_{R})$.
\end{proof}
\end{remark}

\begin{corollary}
Using Proposition~\ref{minmax}, we can rewrite Proposition~\ref{spar} to
\begin{equation}
\log{K}-\log{L}\geq H_{\max}(\rho_{AB}|B)\ .
\end{equation}
This is probably a more intuitive bound, since we analyze state merging from $A$ to $B$.
\end{corollary}

Inequality (\ref{cooop}) is a bound for perfect state merging. Since we want to allow an error $\epsilon$, we need to generalize this to a bound for $\epsilon$-error state merging. To do this we need the following Lemma.

\begin{lemma}
Let $\epsilon\geq0$, $\rho_{AR}\in\mathcal{B}(\mathcal{H}_{A}\otimes\mathcal{H}_{R})$, $\{P_{A}^{y}=\ket{y}\bra{y}_{A}\}_{y\in Y}$ be a projective measurement on $A$ and define $\rho_{AR}'=\sum_{y\in Y}(P_{A}^{y}\otimes\id_{R})\rho_{AR}(P_{A}^{y}\otimes\id_{R})$. Then for every $\sigma'_{AR}\in\mathcal{B}(\mathcal{H}_{A}\otimes\mathcal{H}_{R})$ with $\sigma_{AR}'=\sum_{y\in Y}p_{y}\ket{y}\bra{y}_{A}\otimes\sigma_{R}^{y}$ and $F(\sigma'_{AR},\rho'_{AR})\geq1-\epsilon$,
there exists a $\sigma_{AR}\in\mathcal{B}(\mathcal{H}_{A}\otimes\mathcal{H}_{R})$ with $F(\sigma_{AR},\rho_{AR})\geq1-\epsilon$ and $\sigma'_{AR}=\sum_{y\in Y}(P_{A}^{y}\otimes\id_{R})\sigma_{AR}(P_{A}^{y}\otimes\id_{R})$.
\label{natalie}
\end{lemma}

\begin{proof}
We first prove the statement for $\rho_{AR}$ pure. Define the isometry $U:\ket{y}_{A}\mapsto\ket{y}_{A}\otimes\ket{y}_{Y}$, where $\mathcal{H}_{Y}$ is an ancilla Hilbert space of the same size as $\mathcal{H}_{A}$ and let $\rho'_{ARY}=(U\otimes\id_{R})\rho_{AR}(U^{\dagger}\otimes\id_{R})$. Note that $\rho'_{ARY}$ is pure, i.e.~$\rho'_{ARY}$ is a purification of $\rho_{AR}'$. Now take a $\sigma'_{AR}\in\mathcal{B}(\mathcal{H}_{A}\otimes\mathcal{H}_{R})$ with $\sigma_{AR}'=\sum_{y\in Y}p_{y}\ket{y}\bra{y}_{A}\otimes\sigma_{R}^{y}$ and $F(\sigma_{AR},\rho_{AR})\geq1-\epsilon$. Uhlmann's theorem~\cite{uhlmann:76, jozsa:94} gives us that
\begin{equation}
F(\rho_{AR}',\sigma_{AR}')=\max_{\overline{\sigma}_{ARY}'}F(\rho'_{ARY},\overline{\sigma}_{ARY}')\ ,
\label{678n}
\end{equation}
where the maximization is over all purifications $\overline{\sigma}_{ARY}'$ of $\sigma_{AR}'$. Denote the projector onto $\mathrm{span}(\{\ket{y}_{A}\otimes\ket{y}_{Y}\}_{y\in Y})$ by $Q_{AY}$. Since
\begin{equation}
\begin{split}
F(\rho'_{ARY},\overline{\sigma}_{ARY}')& =F((Q_{AY}\otimes\id_{R})\rho'_{ARY}(Q_{AY}\otimes\id_{R}),\overline{\sigma}_{ARY}')\\
& =F(\rho'_{ARY},(Q_{AY}\otimes\id_{R})\overline{\sigma}_{ARY}'(Q_{AY}\otimes\id_{R}))\ ,
\end{split}
\end{equation}
it is sufficient to maximize in (\ref{678n}) over purifications that lie in the image of $U\otimes\id_{R}$. Denote the state for which the maximum in (\ref{678n}) is taken by $\sigma_{ARY}'$. Since all isometries are injective we can define the inverse of $U\otimes\id_{R}$ on the image of $U\otimes\id_{R}$ and hence $\sigma_{AR}=(U^{-1}\otimes\id_{R})\sigma_{ARY}'((U^{-1})^{\dagger}\otimes\id_{R})\in\mathcal{B}(\mathcal{H}_{A}\otimes\mathcal{H}_{R})$ is well defined. Now this is the $\sigma_{AR}$ we are looking for, since Lemma~\ref{uhhh} gives us that
\begin{equation}
\begin{split}
F(\rho_{AR},\sigma_{AR})& =F((U^{-1}\otimes\id_{R})\rho_{ARY}'((U^{-1})^{\dagger}\otimes\id_{R}),(U^{-1}\otimes\id_{R})\sigma_{ARY}'((U^{-1})^{\dagger}\otimes\id_{R}))\\
& =F(\rho_{ARY}',\sigma_{ARY}')=F(\rho_{AR}',\sigma_{AR}')\geq1-\epsilon\ .
\end{split}
\end{equation}
If $\rho_{AR}$ is not pure, we purify it. This gives us a pure state $\rho_{ARC}$, for which we can go through the same argumentation as above. Since the partial trace is a CPTP map, Lemma~\ref{uhhh} is sufficient to conclude the proof.
\end{proof}

A trace distance version of this Lemma is as follows.

\begin{corollary}
Let $\epsilon\geq0$, $\rho_{AR}\in\mathcal{B}(\mathcal{H}_{A}\otimes\mathcal{H}_{R})$, $\{P_{A}^{y}=\ket{y}\bra{y}_{A}\}_{y\in Y}$ be a projective measurement on $A$ and define $\rho_{AR}'=\sum_{y\in Y}(P_{A}^{y}\otimes\id_{R})\rho_{AR}(P_{A}^{y}\otimes\id_{R})$. Then for every $\sigma'_{AR}\in\mathcal{B}(\mathcal{H}_{A}\otimes\mathcal{H}_{R})$ with $\sigma_{AR}'=\sum_{y\in Y}p_{y}\ket{y}\bra{y}_{A}\otimes\sigma_{R}^{y}$ and $\|\sigma'_{AR},\rho'_{AR}\|_{1}\leq\epsilon$,
there exists a $\sigma_{AR}\in\mathcal{B}(\mathcal{H}_{A}\otimes\mathcal{H}_{R})$ with $\|\sigma_{AR},\rho_{AR}\|_{1}\leq2\sqrt{\epsilon}$ and $\sigma'_{AR}=\sum_{y\in Y}(P_{A}^{y}\otimes\id_{R})\sigma_{AR}(P_{A}^{y}\otimes\id_{R})$.
\label{natalie2}
\end{corollary}

\begin{proof}
Straightforward using Lemma~\ref{fidtr}.
\end{proof}

\begin{proposition}
Let $\rho_{ABR}\in\mathcal{B}(\mathcal{H}_{A}\otimes\mathcal{H}_{B}\otimes\mathcal{H}_{R})$ with $\rho_{ABR}=\ket{\psi}\bra{\psi}_{ABR}$ and $\epsilon\geq0$. Then it holds for any $\epsilon$-error quantum state merging of $\ket{\psi}_{ABR}$ that
\begin{equation}
\log{K}-\log{L}\geq-H_{\min}^{\sqrt{\epsilon}}(\rho_{AR}|R)\ .
\label{boundy}
\end{equation}
\label{sexyy}
\end{proposition}

\begin{proof}
At the begining we have the state $\ket{\psi}_{ABR}\otimes\ket{\Phi_{K}}_{A_{0}B_{0}}$ and in the end we have a state $\rho'_{A_{1}B_{1}B'BR}$ with $\left\|\rho'_{A_{1}B_{1}B'BR}-\ket{\Phi_{L}}\bra{\Phi_{L}}_{A_{1}B_{1}}\otimes\ket{\psi}\bra{\psi}_{BB'R}\right\|_{1}\leq\epsilon$ and $\rho'_{A_{1}B_{1}B'BR}=\sum_{x}p_{x}\rho^{x}_{A_{1}B_{1}B'BR}$, where the $x$ denote the measurement outcomes of the local operations on $A$ (of any hypothetical state merging protocol). Define $\rho'_{A_{1}RX}=\sum_{x}p_{x}\rho_{A_{1}R}^{x}\otimes\ket{x}\bra{x}$, $\sigma'_{A_{1}RX}=\sum_{x}p_{x}\tau_{A_{1}}\otimes\rho_{R}\otimes\ket{x}\bra{x}$ and $\sigma_{X}=\sum_{x}p_{x}\ket{x}\bra{x}$. Note that $\|\rho'_{A_{1}RX}-\sigma'_{A_{1}RX}\|_{1}\leq\epsilon$. We can get
\begin{equation}
\begin{split}
\log L\stackrel{\mathrm{(i)}}{=}H_{\min}(\tau_{A_{1}}\otimes\rho_{R}|\rho_{R}) & \stackrel{\mathrm{(ii)}}{=}H_{\min}(\sum_{x}p_{x}\tau_{A_{1}}\otimes\rho_{R}\otimes\ket{x}\bra{x}|\rho_{R}\otimes(\sum_{x}p_{x}\ket{x}\bra{x}))\\
& =H_{\min}(\sigma'_{A_{1}RX}|\rho_{R}\otimes\sigma_{X})\stackrel{\mathrm{(iii)}}{\leq}H_{\min}(\sigma_{A_{0}AR}|\rho_{R})\\
& \leq H_{\min}^{\sqrt{\epsilon}}(\tau_{A_{0}}\otimes\rho_{R}|\rho_{R})\leq H_{\min}^{\sqrt{\epsilon}}(\tau_{A_{0}}\otimes\rho_{R}|R)\\
& \stackrel{\mathrm{(iv)}}{\leq}\log K+H_{\min}^{\sqrt{\epsilon}}(\rho_{AR}|R)\ .
\end{split}
\end{equation}

Step (i) holds because of Lemma~\ref{add}. In step (ii) we use the fact that the $\ket{x}$ are mutually orthogonal (argumentation analogue as at the beginning of step (b) in the proof of Proposition~\ref{sexy}). To see that step (iii) is correct, let us first deal with the case when the operation on the register $AR$ is given by an isometry on $A$. Then there is only one measurement outcome $x$ and we can just choose $\sigma_{A_{0}AR}$ as the preimage of $\sigma'_{A_{1}RX}$. Due to the same argumentation as in step (a) in the proof of Proposition~\ref{sexy} the estimation holds. If the operation on the register $AR$ is given by a projective measurement on $A$, we can use Lemma~\ref{natalie} to see that there exists a $\sigma_{A_{0}AR}\in\mathcal{B}(\mathcal{H}_{A_{0}}\otimes\mathcal{H}_{A}\otimes\mathcal{H}_{R})$ with $\|\sigma_{A_{0}AR}-\tau_{A_{0}}\otimes\rho_{R}\|_{1}\leq2\sqrt{\epsilon}$, such that $\sigma'_{A_{1}RX}$ is the post measurement state of $\sigma_{A_{0}AR}$. Then Proposition~\ref{sexy} for the state $\sigma_{A_{0}AR}$ justifies step (iii). Furthermore Lemma~\ref{general} shows that the estimate also holds in the general case (argumentation analogue as in step (c) in the proof of Proposition~\ref{sexy}). Finally step (iv) follows from Lemma~\ref{neusi}.
\end{proof}

\chapter{Conclusions}
\label{concl}
\lhead{Chapter 5. \emph{Conclusions}}

We now want to bring together the results of Chapter 4 and point out their exact meaning. We are interested in quantifying the minimal amount of entanglement needed to achieve $\epsilon$-error state merging of $\rho_{ABR}=\ket{\psi}\bra{\psi}_{ABR}$, i.e.~we try to determine the minimal entanglement cost $\log{K}-\log{L}$, where $\log{K}$ stands for the number of bits of pure entanglement at the beginning of the state merging process and $\log{L}$ for the number of bits of pure entanglement in the end.

In Proposition~\ref{sexyy} we showed a lower bound for the entanglement cost for $\epsilon$-error state merging, namely
\begin{equation}
\log{K}-\log{L}\geq-H_{\min}^{\sqrt{\epsilon}}(\rho_{AR}|R)\ .
\label{f}
\end{equation}
In Corollary~\ref{letzter} we showed that there exists an $\epsilon$-error state merging protocol for an entanglement cost of
\begin{equation}
\log{K}-\log{L}=-H_{\min}^{\epsilon^{2}/64}(\rho_{AR}|R)+4\log(\frac{1}{\epsilon})+12\ .
\label{g}
\end{equation}

This can be summarized as follows.

\begin{proposition}
Let $\rho_{ABR}\in\mathcal{B}(\mathcal{H}_{A}\otimes\mathcal{H}_{B}\otimes\mathcal{H}_{R})$ with $\rho_{ABR}=\ket{\psi}\bra{\psi}_{ABR}$ and $\epsilon>0$. Then the minimal entanglement cost for $\epsilon$-error state merging of $\ket{\psi}_{ABR}$ is quantified by
\begin{equation}
\log{K}-\log{L}=-H_{\min}^{\epsilon'}(\rho_{AR}|R)+O(\log(\frac{1}{\epsilon'}))\ ,
\end{equation}
where $\epsilon'\in[\epsilon^{2}/64,\sqrt{\epsilon}]$ and $O(\log(\frac{1}{\epsilon'}))$ denotes an upper bound in the sense of the O-notation.\footnote{For an introduction into the O-notation and precise definitions see~\cite{knuth:76}.}
\end{proposition}

In this sense the protocol described in Proposition~\ref{das} is optimal and we can conclude that the smooth min-entropy is the entropy measure that quantifies the minimal entanglement cost.

The smooth conditional min-entropy of product states asymptotically converges to the conditional von Neumann entropy (Theorem 3.3.6 in~\cite{renner:05}). Hence the average minimal entanglement cost in the asymptotic limit ($n\to\infty$) and for a vanishing error ($\epsilon'\rightarrow0$ in the above notation) becomes
\begin{equation}
\begin{split}
\underset{n\to\infty}{\lim}(\frac{1}{n}(\log{K}-\log{L}))&=\lim_{\epsilon'\to0}\lim_{n\to\infty}(\frac{1}{n}(-H_{\min}^{\epsilon'}(\rho_{AR}^{\otimes n}|R)+O(\log(\frac{1}{\epsilon'})))\\
&=-S(A|R)=S(A|B)\ .
\end{split}
\end{equation}
This is exactly the asymptotic result of Horodecki et al.~\cite{horodecki:05}.

\begin{remark}
Recently it has been shown that the smooth entropy framework and the information spectrum method~\cite{hayashi:03, bowen:06} are asymptotically equivalent~\cite{datta:08}. This means that our result can be reformulated in terms of spectral entropies in the asymptotic case.
\end{remark}

\addtocontents{toc}{\vspace{2em}}

\appendix

\begin{flushright}

\end{flushright}

\chapter{Miscellaneous Facts}
\label{Appendix}
\lhead{Appendix. \emph{Miscellaneous Facts}}

\section{About quantum information theory}

\begin{lemma}[Schmidt-Decomposition]
Let $\rho_{AB}\in\mathcal{B}(\mathcal{H}_{A}\otimes\mathcal{H}_{B})$ with $\rho_{AB}=\ket{\psi}\bra{\psi}_{AB}$. Then there exist orthonormal states $\ket{i}_{A}\in\mathcal{H}_{A}$ and orthonormal states $\ket{i}_{B}\in\mathcal{H}_{B}$ such that
\begin{equation}
\ket{\psi}_{AB}=\sum_{i}\sqrt{\lambda_{i}}\ket{ii}_{AB}\ ,
\end{equation}
where $\lambda_{i}$ are non-negative real numbers satisfying $\sum_{i}\lambda_{i}=1$ known as Schmidt-coefficients. The number of non-zero $\lambda_{i}$ is called Schmidt-rank.
\label{schmidt}
\end{lemma}

\begin{proof}
See~\cite{nielsen:00} page 109.
\end{proof}

\begin{lemma}[Purification]
Let $\rho_{A}\in\mathcal{B}(\mathcal{H}_{A})$. Then there exists a Hilbert space $\mathcal{H}_{R}$ and $\rho_{AR}\in\mathcal{B}(\mathcal{H}_{A}\otimes\mathcal{H}_{R})$ pure such that $\rho_{A}=\textrm{tr}_{R}(\rho_{AR})$.
\label{purification}
\end{lemma}

\begin{proof}
See~\cite{nielsen:00} page 110.
\end{proof}

\begin{lemma}
Let $\rho_{A}\in\mathcal{B}(\mathcal{H}_{A})$ and let $\left\{M_{x}\right\}_{x\in X}$ be a measurement on $A$. Then there exists a projective measurement $\left\{P_{x}\right\}_{x\in X}$, a Hilbert space $\mathcal{H}_{X}$ with mutually orthogonal basis $\left\{\ket{x}\right\}_{x\in X}$ and a unitary evolution $U$ on $\mathcal{H}_{A}\otimes\mathcal{H}_{X}$ such that
\begin{equation}
\begin{split}
\textrm{tr}_{X}(\sum_{x\in X}(\id\otimes P_{x})U(\rho\otimes\ket{0}\bra{0})U^{\dagger}(\id\otimes P_{x})^{\dagger})&=\textrm{tr}_{X}(\sum_{x\in X}M_{x}(\rho\otimes\ket{x}\bra{x})M_{x}^{\dagger})\\
&=\sum_{x\in X}M_{x}\rho M_{x}^{\dagger}\ .
\end{split}
\end{equation}
\label{general}
\end{lemma}

\begin{proof}
See~\cite{nielsen:00} page 94.
\end{proof}

\begin{lemma}[Stinespring Dilation]
Let $\rho_{A}\in\mathcal{B}(\mathcal{H}_{A})$ and let $\left\{M_{x}\right\}_{x\in X}$ be a measurement. Then there exists a Hilbert space $\mathcal{H}_{B}$ and a unitary evolution $U$ on $\mathcal{H}_{A}\otimes\mathcal{H}_{B}$ such that
\begin{equation}
\sum_{x\in X}M_{x}\rho M_{x}^{\dagger}=\textrm{tr}_{B}(U(\rho\otimes\ket{0}\bra{0})U^{\dagger})\ .
\end{equation}
\label{stine}
\end{lemma}

\begin{proof}
See~\cite{stinespring:55}.
\end{proof}

\section{About some technical stuff}

\begin{lemma}[Commutativity of partial trace with identity]
Let $\rho_{AB}\in\mathcal{B}(\mathcal{H}_{A}\otimes\mathcal{H}_{B})$ and $\sigma_{A}\in\mathcal{B}(\mathcal{H}_{A})$. Then
\begin{equation}
\textrm{tr}_{B}(\rho_{AB}(\sigma_{A}\otimes\id_{B}))=\textrm{tr}_{B}(\rho_{AB})\sigma_{A}\ .
\end{equation}
\label{gross}
\end{lemma}

\begin{proof}
Straightforward.
\end{proof}

\begin{lemma}
Let $S$ be a hermitian operator on $\mathcal{H}$ and $\sigma$ be a nonnegative operator on $\mathcal{H}$. Then
\begin{equation}
\left\|S\right\|_{1}\leq\sqrt{\textrm{tr}(\sigma)}\left\|\sigma^{-1/4}S\sigma^{-1/4}\right\|_{2}\ .
\end{equation}
\label{5.1.3}
\end{lemma}

\begin{proof}
The above statement can be rewritten to
\begin{equation}
\left\|S\right\|_{1}\leq\sqrt{\textrm{tr}(\sigma)\textrm{tr}(S\sigma^{-1/2}S\sigma^{-1/2})}\ .
\end{equation}
This is Lemma 5.1.3 in~\cite{renner:05}.
\end{proof}

\begin{lemma}
Let $\rho$, $\sigma\in\mathcal{B}(\mathcal{H})$ such that $\sigma$ is invertible. Then the operator $\lambda\cdot\sigma-\rho$ is nonnegative if and only if
\begin{equation}
\lambda_{\max}(\sigma^{-1/2}\rho\sigma^{-1/2})\leq\lambda\ .
\end{equation}
\label{ja}
\end{lemma}

\begin{proof}
This is a special case of Lemma B.5.3 in~\cite{renner:05}.
\end{proof}

\addtocontents{toc}{\vspace{2em}}

\lhead{\emph{Bibliography}}
\bibliographystyle{abbrv}

\end{document}